\documentclass[a4paper,twocolumn,prb,showpacs]{revtex4}
\usepackage{amsmath,amsfonts,amssymb}
\usepackage{amsfonts}
\usepackage{graphicx}
\usepackage{mathbbol}

\begin{document}

\title{Controllable Goos-H\"{a}nchen shifts and spin beam splitter for
ballistic electrons in a parabolic quantum well under a uniform magnetic field}

\author{Xi Chen$^{1,2}$}
\email[Author to whom correspondence should be addressed. ] {xchen@shu.edu.cn}

\author{Xiao-Jing Lu$^{1}$}
\author{Yan Wang$^{1}$}
\author{Chun-Fang Li$^{1,3}$}

\affiliation{$^{1}$Department of Physics, Shanghai University,
200444 Shanghai, China}
\affiliation{$^{2}$Departamento de Qu\'{\i}mica-F\'{\i}sica,
UPV-EHU, Apdo 644, 48080 Bilbao, Spain}
\affiliation{$^{3}$State Key Laboratory of Transient Optics and Photonics,
Xi'an Institute of Optics and Precision Mechanics of CAS, 710119 Xi'an, China}

\date{\today}

\begin{abstract}
The quantum Goos-H\"{a}nchen shift for ballistic electrons
is investigated in a parabolic potential well
under a uniform vertical magnetic field.
It is found that the Goos-H\"{a}nchen shift can be negative as well as positive, and
becomes zero at transmission resonances. The beam shift depends not only on the incident energy and incidence angle,
but also on the magnetic field and Landau quantum number.
Based on these phenomena, we propose an alternative way to
realize the spin beam splitter in the proposed spintronic device, which can completely separate spin-up
and spin-down electron beams by negative and positive Goos-H\"{a}nchen shifts.
\end{abstract}


\pacs{72.25.Dc, 42.25.Gy, 73.21.Fg, 73.23.Ad}                    

\maketitle


\section{Introduction}

The longitudinal Goos-H\"{a}nchen shift is well known for a light beam totally reflected from an
interface between two dielectric media \cite{Lotsch}.
This phenomenon, suggested by Sir Isaac Newton,
was observed firstly in microwave experiment by Goos and H\"{a}nchen \cite{Goos}
and theoretically explained by Artmann in term of stationary phase method
\cite{Artmann}. Up till now, the investigations of the Goos-H\"{a}nchen shift
have been extended to different areas of physics \cite{Lotsch}, such as
quantum mechanics \cite{Renard,Carter}, acoustics \cite{Briers}, neutron
physics \cite{Ignatovich,Haan}, spintronics \cite{Chen-PRB}, atom optics
\cite{Zhang-WP}, and graphene \cite{Zhao,Beenakker-PRL,Sharma}, based on the particle-wave duality
in quantum mechanics.

Historically, the quantum Goos-H\"{a}nchen shifts and relevant transverse Imbert-Fedorov shifts
have been once studied for relativistic Dirac electrons in 1970s \cite{Miller,Fradkin}.
With the development of the semiconductor technology, the Goos-H\"{a}nchen shift for ballistic electrons
in two-dimensional electron gas (2DEG) system becomes one of the important subjects
on the ballistic electron wave optics
\cite{Wilson-G-G,Sinitsyn,XChen-PLA,XChen-JAP}. It was found that the
lateral shifts of ballistic electrons transmitted through
a semiconductor quantum
barrier or well can be enhanced by transmission resonances, and become negative as well as
positive \cite{XChen-JAP}. The interesting Goos-H\"{a}nchen shift, depending on the spin polarization,
could also provide an alternative way to realize the spin filter and spin beam splitter
in spintronics \cite{Chen-PRB},
in the same way as other optical-like phenomena
including double refraction \cite{Ramaglia} and negative refraction \cite{Zhang}
in spintronics optics \cite{Khodas}.

In this paper, we will investigate the Goos-H\"{a}nchen shifts for
ballistic electrons in a parabolic quantum well under a uniform magnetic field,
in which the high spin polarization and electron transmission probability can be achieved \cite{Wan}.
It is shown that the lateral shift can be negative as well as positive. As a matter of fact,
the Goos-H\"{a}nchen shift discussed here is different
from that in the magnetic-electric nanostructure \cite{Chen-PRB},
where the $\delta$ magnetic field is considered.
In such quantum well, the uniform magnetic field bends the trajectory of electron continuously,
so that electrons exhibit cyclotron motion. Then this implies that the behavior of electrons
in such system has no direct analogy with linear propagation of light \cite{Sharma}. Meanwhile,
it is also suggested that Landau quantum number will have great effect on the Goos-H\"{a}nchen shift.
We consider the Goos-H\"{a}nchen shift for ballistic electrons in a parabolic quantum well under a uniform magnetic field
and its dependence on the magnetic field and Landau energy level, which to our knowledge,
has not been investigated so far. More importantly, the Goos-H\"{a}nchen shift for ballistic electrons
depends on the spin polarization by Zeeman interaction, which is similar to that for neutrons \cite{Haan}.
The realization of the negative and positive Goos-H\"{a}nchen shifts corresponding to spin-up and spin-down polarized electrons
may have future applications in the proposed spintronic devices, such as spin filter and
spin beam splitter.

\section{Theoretical model}
\label{model}

\begin{figure}[t]
\scalebox{0.60}[0.60]{\includegraphics{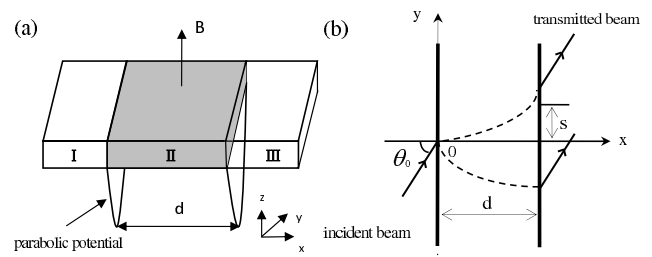}}
 \caption{ (a) Schematic diagram for 2DEG
with the parabolic quantum well under a uniform magnetic field.
(b) Negative and positive Goos-H\"{a}nchen shifts of ballistic electrons are presented in this configuration.}\label{Fig.1}
\end{figure}

We consider a 2DEG structure, with a confining potential in
the central region, as shown in Fig. 1 (a). The confinement potential
is assumed to be parabolic in the transverse $y$ direction and the
electron transports in the 2DEG occur in the plane of $x$-$y$. 
A uniform magnetic $B$ field is applied along the perpendicular $z$
direction and limited to within the parabolic confinement region. In
practice, such a $B$-field configuration can be achieved by means of a
ferromagnetic gate stripe on top of the 2EDG heterostructure \cite{Wan}. The
Hamiltonian within the parabolic quantum well is then given by
\begin{equation}
H=(1/2m)(\textbf{p}-\textbf{A})^{2}+H_{conf}+H_{z}
\end{equation}
where $m$ is the effective mass of the electron and $\mathbf{p}$ is
the momentum of the electron. The Landau gauge is chosen for the
magnetic vector potential, i.e., $\textbf{A}=(By,0,0)$, where $B$ is
the magnetic field strength. The parabolic confinement energy can be
expressed as $H_{conf}=\frac{1}{2}m \omega_{0}^{2}y^{2}$, while the
Zeeman interaction term is given by $
H_{z}=\frac{1}{2}\mu_{B}g\sigma B$, so $\mu_{B}$ is the Bohr
magneton, $g$ is Land\'{e} factor and $\sigma=\pm 1$ denotes the spin
orientation parallel or antiparallel to the reference $z$ axis. In this system,
the wave functions of plane wave for electrons in three regions can be expressed as
\begin{eqnarray}
\Psi_{I} (x,y) &=& (e^{ik_{x}x}+Ae^{-ik_{x}x})e^{ik_{y}y}, ~~~~~~~~x<0
\\
\Psi_{II} (x,y)&=& (Be^{ik'_{x} x}+Ce^{-ik'_{x} x})\psi_{2}(y), ~ 0<x<d
\\
\Psi_{III} (x,y)&=& D e^{ik_{x}(x-d)}e^{ik_{y}y}, ~~~~~~~~~~~~~~~~x>d
\end{eqnarray}
where $k_{x}=\sqrt{2 m E/ \hbar^2 -k_{y}^{2}}$ is the traveling wave vector in 2DEG.
The Schr\"{o}dinger equation in the parabolic potential region can
be written in the form of a harmonic oscillator centered at
$\widetilde{Y}$ with angular frequency $\widetilde{\omega}_{c}$,
\begin{eqnarray}
 \nonumber
&& \left[\frac{m \widetilde{\omega}_{c}^{2}}{2}(y-\widetilde{Y})^{2}+\frac{m \omega_{0}^{2} \omega_{c}^{2}}{2\widetilde{\omega}_{c}^{2}}Y^{2}+\frac{\mu_{B}g\sigma}{2}
B-\frac{\hbar^{2}}{2m}\frac{d^{2}}{dy^{2}}\right] \Psi_{II} \\
 &&~~~~~~~~~~~~~~~~~~~~~~~~~~~~~~~~~~~~~~~~~~~~~~~~~~~~~= E \Psi_{II}.
\end{eqnarray}
with
$\widetilde{Y}=(\omega_{c}/\widetilde{\omega}_{c})^{2} Y $,
$\widetilde{\omega}_{c}^{2}=\omega_{0}^{2}+\omega_{c}^{2}$, $\omega_{c}=B e/m$, $Y=\hbar
k'_{x}/eB$, $\omega_{c}$ is the cyclotron frequency. So the solution is given by a linear combination of Hermite
polynomials as follows:
\begin{equation}
\psi_2 (y) = \left(\frac{1}{2^n n! \gamma \sqrt{\pi}} \right)^{1/2} e^{-[(y-\widetilde{Y})^2/2 \gamma^2]} H_n \left(\frac{y-\widetilde{Y}}{\gamma}\right),
\end{equation}
where $ \gamma^2=\hbar/\sqrt{e^2 B^2 + m^2 \omega^2_0}$. The corresponding eigenenergy is
\begin{equation}
E_{n}=(n+\frac{1}{2})\hbar\widetilde{\omega}_{c}+\frac{m \omega_{0}^{2} \omega_{c}^{2}}{2\widetilde{\omega}_{c}^{2}}Y^{2}+\frac{1}{2}\mu_{B}g\sigma
B.
\end{equation}
The eigenstates of the system form a set of Landau-like states with
energy splitting proportional to $B$. Within the central potential
region, the longitudinal wave vector $k'_{x}$ is given by
\begin{equation}
\label{dispersion}
k'_{x}=\sqrt{\frac{2m\widetilde{\omega}_{c}^{2}[E_{n}-(n+\frac{1}{2})\hbar\widetilde{\omega}_{c}-\frac{1}{2}\mu_{B}g\sigma
B]}{\hbar^{2}\omega_{0}^{2}}},
\end{equation}
which is spin-dependent due to Zeeman interaction.
Interestingly, when the only plane wave is considered, the spatial location of the eigenstates inside the
region of potential well is around $Y=\hbar
k'_{x}/eB$, which can be rewritten by
\begin{equation}
Y= \nu (n, k'_x) \frac{\omega_{0}^{2}+\omega_{c}^{2}}{\omega_{c} \omega_{0}^{2} },
\end{equation}
with the velocity
\begin{equation}
\nu (n, k'_x) = \frac{1}{\hbar} \frac{\partial E_n}{\partial k'_x}= \frac{\omega_{0}^{2} }{\widetilde{\omega}_{c}^{2}} \frac{\hbar k'_x}{m} .
\end{equation}
The transverse location for each plane wave eigenstate is proportional to the velocity and magnetic field. From the classical viewpoint,
the spatial shift can be reasonably explained by Lorentz force \cite{Datta}. As a consequence,
the transverse shifts for the forward and backward propagating states inside the central region are positive and negative,
since Lorentz force is opposite for electrons moving in the opposite direction.
However, the lateral shift predicted by Goos-H\"{a}nchen effects will be totally different.
In what follows, we will discuss the spin-dependent Goos-H\"{a}nchen shift of electron beam, instead of plane wave, in such a configuration.


\section{Goos-H\"{a}nchen shifts and spin beam splitter}

\subsection{Stationary phase method}

When the finite-sized incident electron beam is considered, the wave function of incident beam
can be assumed to be
\begin{equation}
\label{incident beam}
\Psi_{in} (x, y) = \frac{1}{\sqrt{2\pi}}\int A(k_y-k_{y0}) e^{i (k_{x}x + k_{y}y)} d k_y,
\end{equation}
with the angular spectrum distribution $A(k_y-k_{y0})$ around the central wave vector $k_{y0}$, then the transmitted beam can be expressed by
\begin{equation}
\label{transmitted beam}
\Psi_{tr} (x, y) = \frac{1}{\sqrt{2\pi}} \int D A(k_y-k_{y0}) e^{i [k_{x}(x-d)+ k_{y}y ]} d k_y,
\end{equation}
where the transmission coefficient $D=\exp{(i \phi)/g}$ is determined by the boundary conditions at $x=0$ and $x=d$ with
\begin{equation}
g \exp{(i \phi)= \cos{(k'_{x} d)} + i \left(\frac{k^2_{x}+k^{'2}_{x}}{2 k_{x} k'_{x}}\right) \sin{(k'_{x} d})},
\end{equation}
which leads to the phase shift $\phi$ in term of
\begin{equation}
\tan \phi= \frac{k_{x}^{2}+k^{'2}_{x}}{2k_{x}k'_{x}}\tan(k_{x}^{'}d).
\end{equation}

To find the position where $\Psi_{tr} (x, y)$ is maximum, that is, the lateral shift of transmitted beam,
we look for the place where the phase of transmitted beam $\Phi= k_x (x-d) + k_y y + \phi$, has an extremum when differentiated with respect to
$k_y$, ie, $\partial \Phi / \partial k_y =0$ \cite{Bohm}. So according to the stationary phase approximation,
the Goos-H\"{a}nchen shift for ballistic electrons at $x=d$
is defined as \cite{Chen-PRB}
\begin{equation}
\label{definition}
s=-\frac{d\phi}{dk_{y0}}.
\end{equation}
It is noted that the subscript $0$ denotes the value at $k_y=k_{y0}$, namely $\theta= \theta_0$. Thus,
the Goos-H\"{a}nchen shift, as described in Fig. 1 (b), can be obtained by
\begin{equation}
\label{Goos-Hanchen shift}
s=\frac{d \tan\theta_0}{2 g^2_0 }\left(1-\frac{k^{'2}_{x0}}{k_{x0}^{2}}\right)\frac{\sin{(2 k'_{x0}d)}}{2k'_{x0
}d},
\end{equation}
where $\theta_0$ is the incidence angle and $\tan \theta_0=k_{y0}/k_{x0}$.
Obviously, it is clearly seen from Eq. (\ref{Goos-Hanchen shift}) that
the Goos-H\"{a}nchen shifts are negative as well as positive,
depending on the $k^{'2}_{x0}/k^2_{x0}$ and $\sin{(2 k'_{x0}d)}$.

\begin{figure}[t]
\begin{center}
\scalebox{0.32}[0.35]{\includegraphics{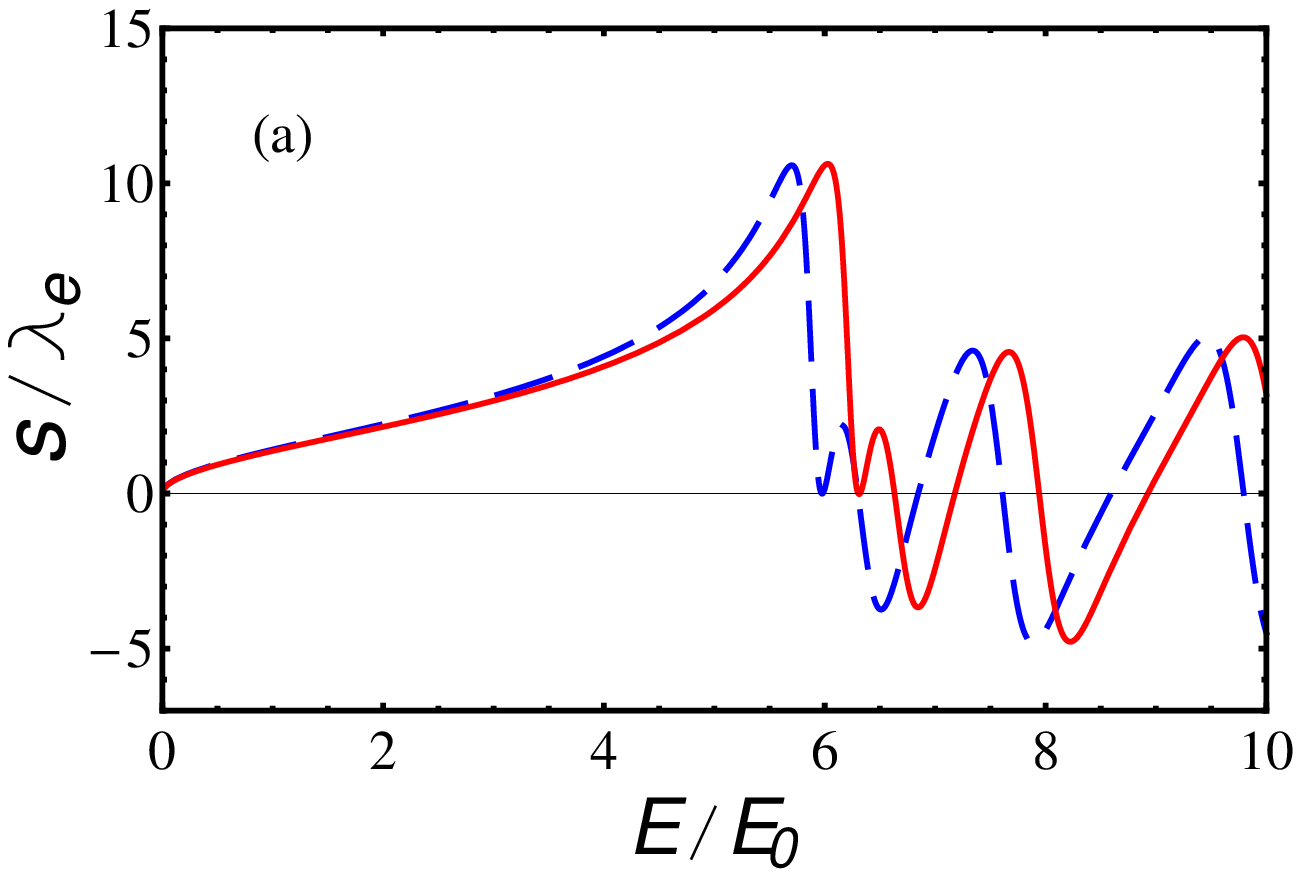}}
\scalebox{0.32}[0.35]{\includegraphics{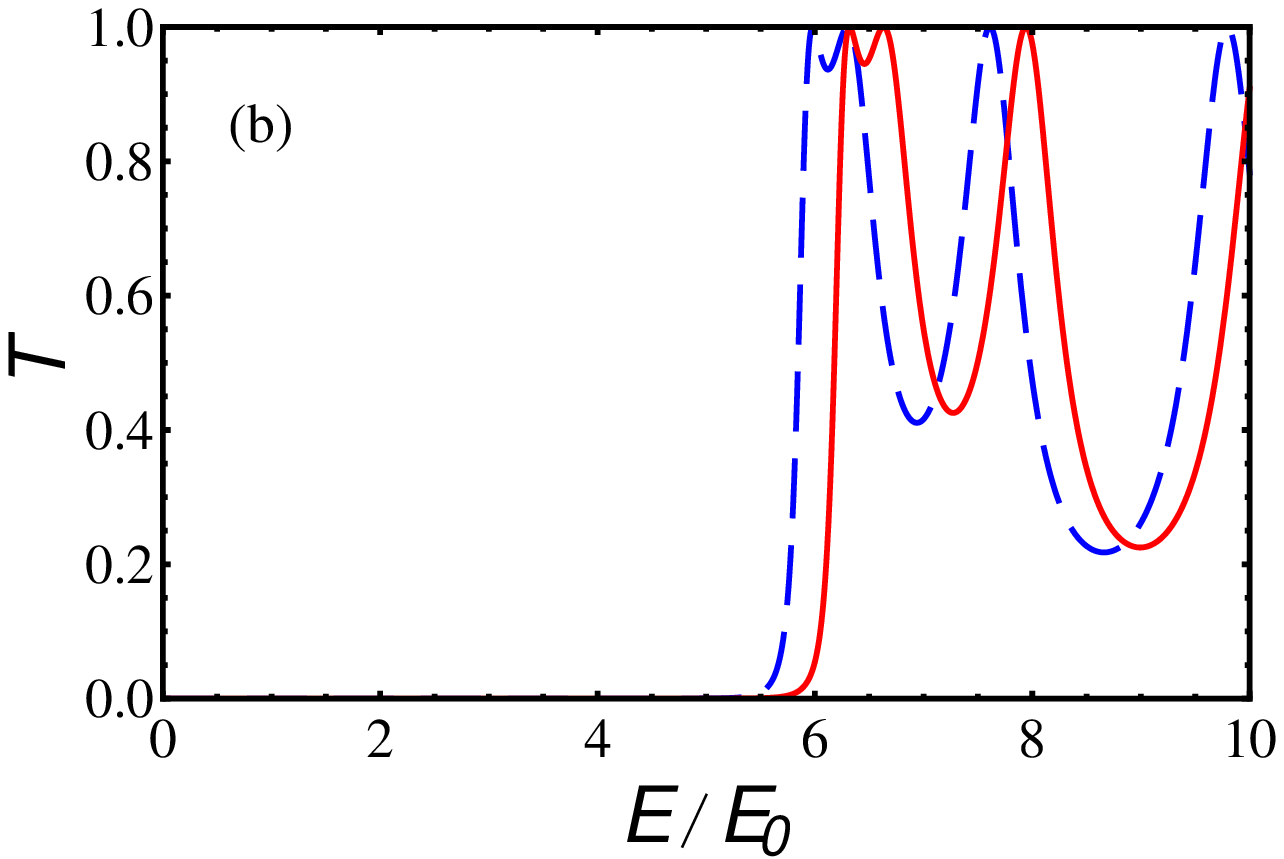}}
 \caption{(Color online) Dependence of Goos-H\"{a}nchen shift (a) and transmission probability (b)
on the incident energy $E/E_0$, where $\theta_0=70^{\circ}$ and the other physical parameters are respectively
$m=0.013m_{e}$, $g=51$, $d=50$ nm, $B=0.5$ T and $n=5$. Solid (red) and dashed (blue) lines correspond to
spin-up and spin-down polarized electrons.} \label{Fig.2}
\end{center}
\end{figure}

In the propagating case, when the transmission resonances $k'_{x0} d=m \pi ~(m=1,2,3...)$ or anti-resonances $k'_{x0} d=(m+1/2) \pi ~(m=1,2,3...)$
occur, the Goos-H\"{a}nchen shift is zero, which means the
positions in the $y$ direction are the same for both incident and transmitted electrons.
As a matter of fact, when measured with reference to the geometrical
prediction from electron optics, the ``zero" lateral shifts will become essentially negative values.
Whereas, when the incident energy is less than the critical energy,
\begin{equation}
\label{condition}
E < E_{c}=(n+\frac{1}{2})\hbar\widetilde{w}_{c}+\frac{1}{2}\mu_{B}g\sigma B,
\end{equation}
$k'_{x0}$ becomes imaginary, thereby $k^{'2}_{x0}/k^2_{x0}<0$. Substituting  $k'_{x0}= i \kappa_0$ into  Eq. (\ref{Goos-Hanchen shift}),
we end up with the following expression:
\begin{equation}
\label{Goos-Hanchen shift-evanscent case}
s=\frac{d \tan\theta_0}{2 g'^2_0 }\left(1+\frac{\kappa^2_{0}}{k_{x0}^{2}}\right)\frac{\sinh{(2 \kappa_{0}d)}}{2 \kappa_{0} d},
\end{equation}
with
\begin{equation}
g^{'2}_{0} = \cosh^2{(\kappa_0 d)} +\left(\frac{k^2_{x}- \kappa^{2}_0}{2 k_{x} \kappa_0 }\right)^2 \sinh^2{(\kappa_0 d}).
\end{equation}
In the evanescent case, the Goos-H\"{a}nchen shifts are always positive,
which seems similar to the Goos-H\"{a}nchen shifts in a single semiconductor barrier \cite{XChen-PLA}.
In the opaque limit, $\kappa_0 d \gg 1$, the lateral shift saturates to
\begin{equation}
s= \frac{2 \kappa_0 \tan\theta_0}{k_{x0}^{2} + \kappa^{2}_0} \simeq  \frac{2 \tan\theta_0}{\kappa_0},
\end{equation}
which is independent of the width $d$ of potential well.
In the following discussions, we will study the Goos-H\"{a}nchen shifts for different polarized electrons and their modulation by the
magnetic field.

First of all, an example of the Goos-H\"{a}nchen shifts (in the units of $\lambda_{e}=2 \pi \hbar/\sqrt{2m_{e}E}$)
for ballistic electrons as function of incident energy $E/E_0$ in the quantum well
under the uniform vertical magnetic field is displayed in Fig. \ref{Fig.2} (a), where the quantum well
is made from InSb semiconductor, the physical parameters are respectively
$m=0.013m_{e}$, $m_{e}$ is the bare electron mass, $g=51$,
the length of potential well $d=50$ nm, the parabolic well of depth, $\hbar \omega_{0}= 2$ meV,
the applied magnetic field $B=0.5$ T, $n=5$, and $E_{0}=\hbar \omega_{c}= 4.45$ meV.
It is clearly seen from the condition (\ref{condition}) that the critical energies
are different for spin-up and spin-down polarized electrons, which correspond to
$E^{+}_c = 26.11$ meV and $E^{-}_c = 27.58$ meV, respectively, for the parameters used here.
As shown in Fig. \ref{Fig.2} (a), the behavior of Goos-H\"{a}nchen shifts coincides with the theoretical analysis mentioned above,
that is to say, the Goos-H\"{a}nchen shifts are always positive when the incident energy is less
than the critical energy $E_c$, while the lateral shifts become negative as well as positive periodically related to the
transmission resonances. In addition, Fig. \ref{Fig.2} (b) also shows the dependence of transmission probability $T=|D|^2$
on the incident energy.
This quantity can be connected with the measurable ballistic conductance $G$, according to the well-known Landauer-B\"{u}ttiker formula
at zero or non-zero temperature \cite{Datta,Buttiker}.

\begin{figure}[t]
\scalebox{0.32}[0.35]{\includegraphics{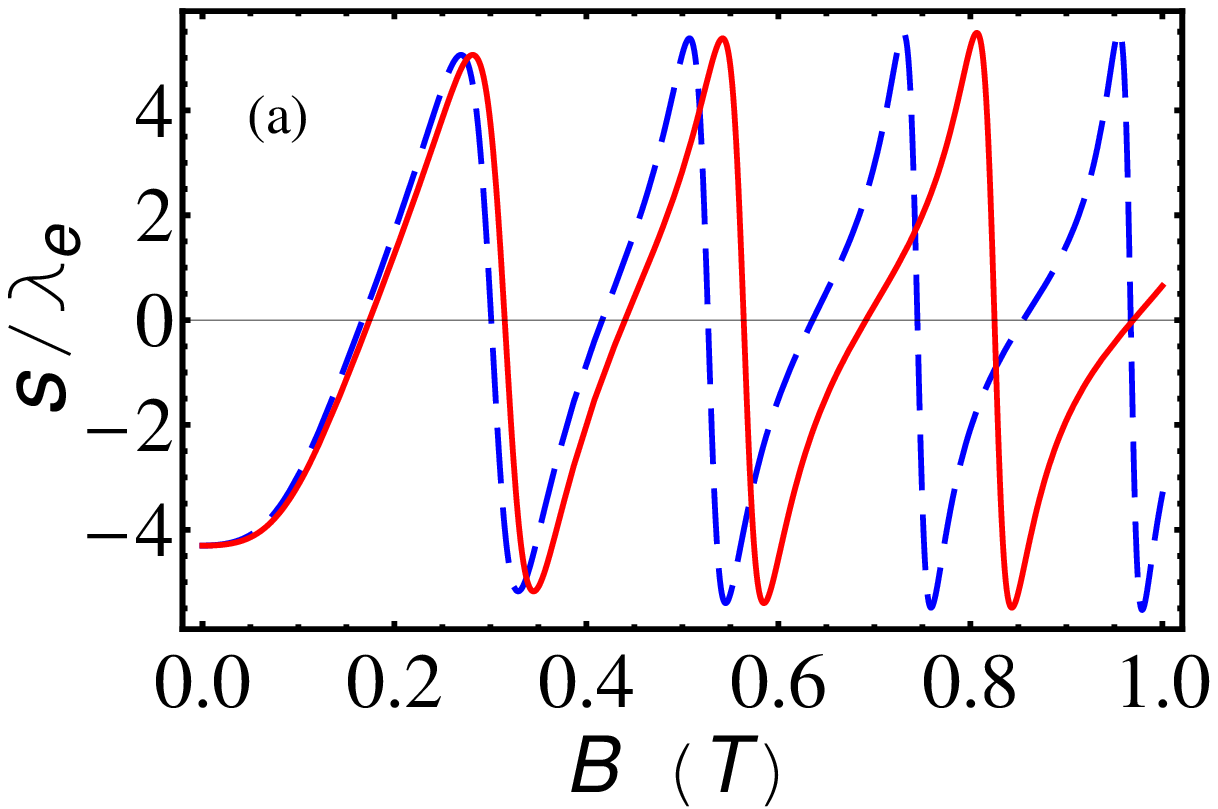}}
\scalebox{0.32}[0.35]{\includegraphics{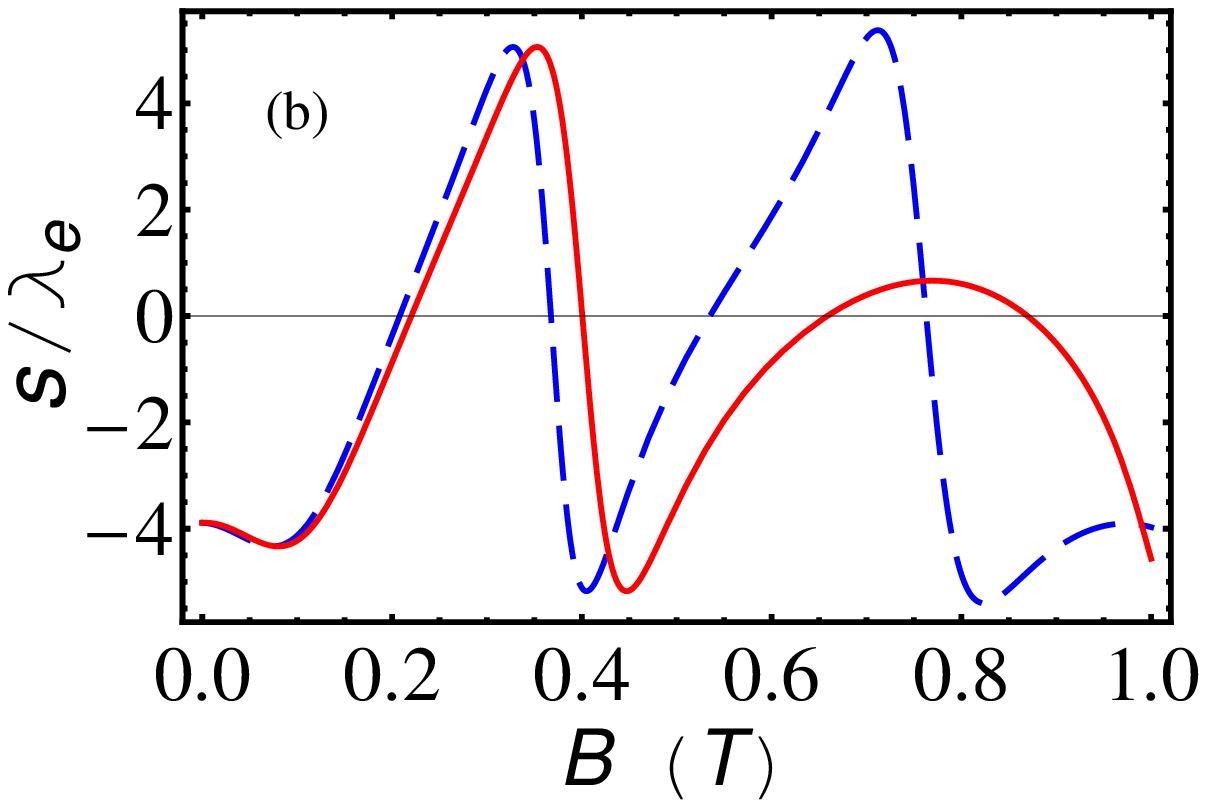}}
\scalebox{0.32}[0.35]{\includegraphics{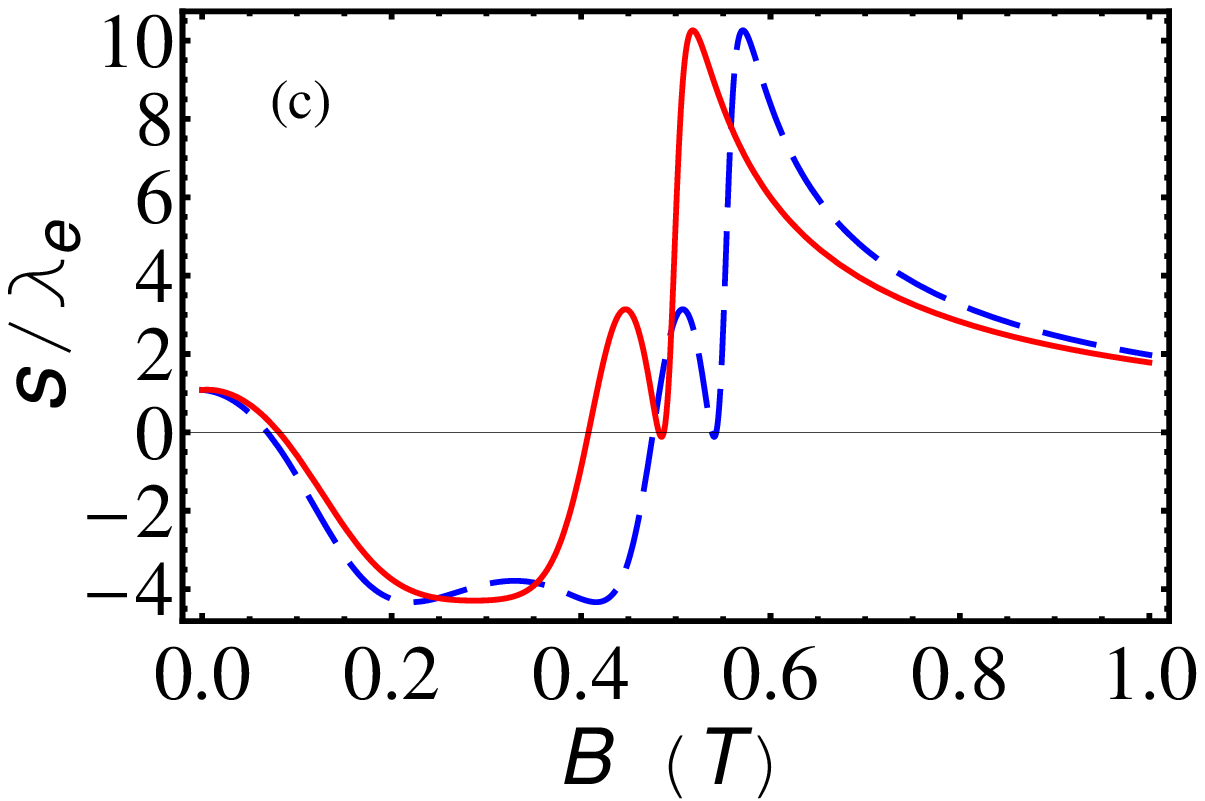}}
\scalebox{0.32}[0.35]{\includegraphics{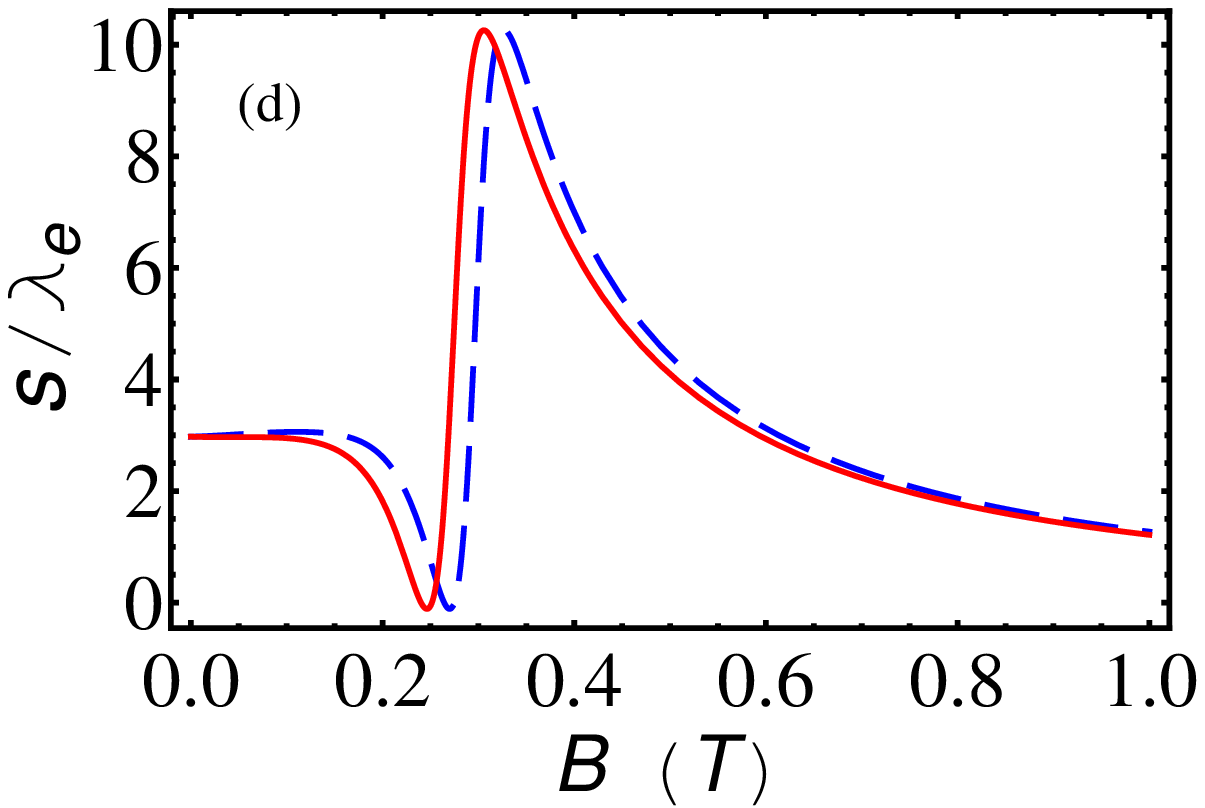}}
 \caption{(Color online) Dependence of Goos-H\"{a}nchen shift on the strength of  magnetic field $B$ with different Landau quantum number $n=0$ (a), $n=1$ (b), $n=3$ (c),
 $n=5$ (d),
where $\theta_0=70^{\circ}$, $E=17.81$ meV, and the other physical parameters are the same as those in Fig. \ref{Fig.2}.
Solid (red) and dashed (blue) lines correspond to
spin-up and spin-down polarized electrons.} \label{Fig.3}
\end{figure}

More interestingly, the spin-up and spin-down polarized electron beams
can be separated by spatial shifts, due to their energy dispersion relation depending on the polarization.
Based on the properties of Goos-H\"{a}nchen shifts, the simplest way to realize the energy filter
by the Goos-H\"{a}nchen shifts is as follow. We can choose the incident energy within the range
of $E^{+}_{c}<E<E^{-}_{c}$. This suggests that the spin-down polarized electrons for $E>E^{-}_{c}$ can traverse through the structure
in the propagating mode with high transmission probability, while the spin-up polarized electrons for $E< E^{+}_{c}$ tunnel through it in the evanescent mode with very low
transmission probability, as described in Fig. \ref{Fig.2} (b). So this provides an alternative way to design a spin spatial filter
with energy width $\Delta E =\mu_{B} g B = 1.46$ meV for the parameters in Fig. \ref{Fig.2}.

Next, we will discuss the influence of magnetic field and Landau energy level
on the Goos-H\"{a}nchen shifts in such semiconductor device. Figure \ref{Fig.3}
illustrates the dependence of Goos-H\"{a}nchen shifts on the strength of magnetic field
$B$ with different Landau quantum number $n$, where $n=0$ (a), $n=1$ (b), $n=3$ (c),
 $n=5$ (d), $\theta_0=70^{\circ}$, $E=17.81$ meV, and the other physical parameters are the same as those in Fig. \ref{Fig.2}.
Solid (red) and dashed (blue) lines correspond to spin-up and spin-down polarized electrons. Basically,
the behavior of Goos-H\"{a}nchen shifts is different from that of transverse shifts for the plane wave, as described in Sec. \ref{model}.
According to the definition of critical energy $E_c$, under the low strength of magnetic field, the incident energy
will be larger than the critical energy, thus the electron can propagate though the quantum well
with negative and positive Goos-H\"{a}nchen shifts, which can be adjusted by the conditions for
transmission resonances with changing magnetic fields. In other word,
in the propagating case, the Goos-H\"{a}nchen shifts can be changed from negative to positive by controlling the strength of magnetic field,
vice versa. However, the Goos-H\"{a}nchen shifts finally become positive with increasing the strength of magnetic field, due to the fact that
the critical energy becomes larger than the incident energy with enough large magnetic field $B_c$, and the propagation of electrons is actually evanescent in this case.
To understand the dependence of shifts on the magnetic field better, we have to calculate the critical magnetic field $B_c$
from Eq. (\ref{condition}) by solving the following equation,
$(n+1/2)\hbar e^2 B^2_c / m^2 + \mu_{B}g\sigma B_c/2+ (n+1/2)\hbar \omega^2_0-E=0$,
for the fixed incident energy $E$. For example, the critical magnetic fields for spin-up and spin-down polarized electrons
can be numerically obtained as follows: $B^{+}_c =1.02$ T and $B^{-}_c = 1.48$ T for $n=1$,
$B^{+}_c =0.49$ T and $B^{-}_c = 0.55$ T for $n=3$, $B^{+}_c =0.28$ T and $B^{-}_c = 0.30$ T for $n=5$.
So these results can be applicable to explain the most striking effect around the critical magnetic field $B_c$ in Fig. \ref{Fig.3} (b)-(d),
while the similar behavior is not shown in Fig. \ref{Fig.3} (a),
because the critical magnetic fields, $B^{+}_c =3.00$ T and $B^{-}_c = 5.98$ T for $n=0$, are too large,
we just show the range of magnetic field from $0$ to $1$ T due to the
physical restriction to applied magnetic field in the laboratory. In addition, another observation from Fig. \ref{Fig.3} is that,
with increasing Landau quantum number $n$, the curves of Goos-H\"{a}nchen shifts move leftwards.
That is to say, for the given incident energy, Landau energy level
leads to the fact that the resonant peak or the critical strength of magnetic field $B_c$ shifts
towards the low magnetic field strength region as increasing
Landau quantum number $n$. Compared to the previous results in the magnetic-electric nanostructure \cite{Chen-PRB},
the magnetic field and Landau quantum number provide more freedom to control the quantum Goos-H\"{a}nchen shift for ballistic electrons,
which is useful to its application in the semiconductor devices.

\begin{figure}[t]
\scalebox{0.35}[0.38]{\includegraphics{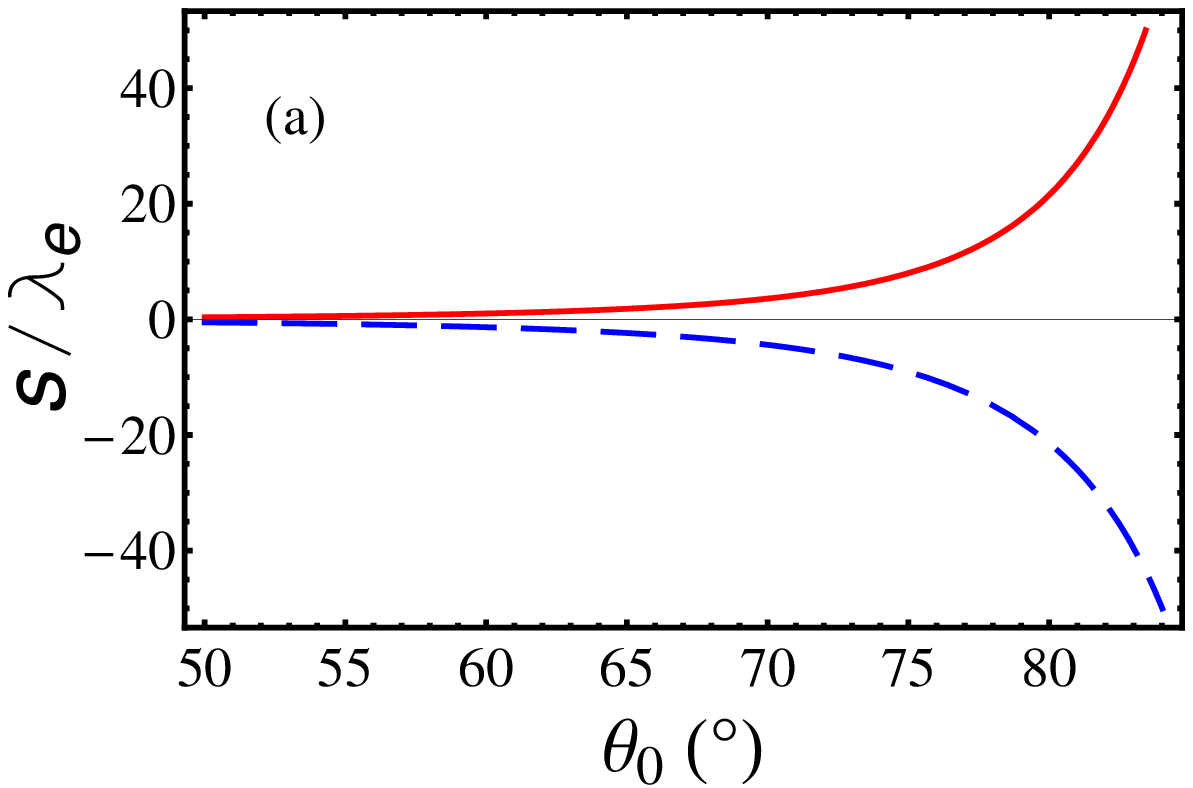}}
\scalebox{0.35}[0.38]{\includegraphics{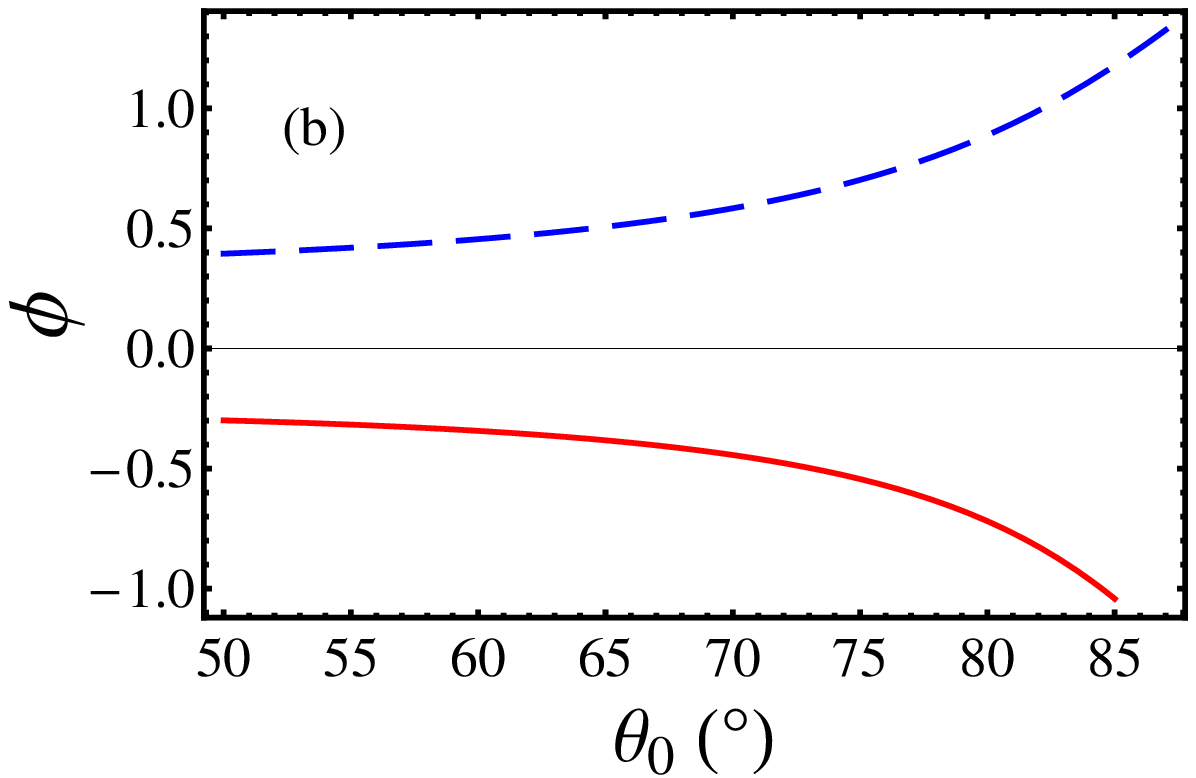}}
 \caption{(Color online) Spin-dependent Goos-H\"{a}nchen shifts (a) and corresponding phase shifts (b)
 as the function of incidence angle, where $E/E_0 = 7.8$ and the other physical parameters are
 the same as those in Fig. \ref{Fig.2}. Solid (red) and dashed (blue) lines represent spin-up and spin-down
 polarized electrons, respectively.} \label{Fig.4}
\end{figure}

Now, we will discuss the spin beam splitting by the spin-dependent Goos-H\"{a}nchen shifts
at various incidence angles. In general, the Goos-H\"{a}nchen shifts can also be modulated by the incidence angles, due to the energy dispersion (\ref{dispersion}).
In Fig. \ref{Fig.4} (a), we would like to emphasize that the spin-up and spin-down polarized electrons can be spatially separated by
the spin-dependent Goos-H\"{a}nchen shifts at large incidence angles, for example $\theta_0 = 80^\circ$, where incident energy $E/E_0 = 7.8$, and
the other physical parameters are the same as those in Fig. \ref{Fig.2}.
To confirm this intriguing result, the phase shifts of transmitted electrons have also
been investigated in Fig. \ref{Fig.4} (b). In detail, the negative slope of the phase shift with respect to
the incidence angle will result in the positive Goos-H\"{a}nchen shift, whereas the positive one will
lead to the negative Goos-H\"{a}nchen shift, which are in agreement with theoretical predictions by stationary phase method.
Thus, the Goos-H\"{a}nchen shifts can be explained by
the reshaping process of the transmitted beam, since each plane wave component undergoes the
different phase shift due to the multiple reflections inside the quantum well \cite{XChen-PLA}.
The negative and positive beam shifts are usually considered as the consequence of constructive and
destructive interferences between the plane wave components of electron beam.

\subsection{Numerical simulation}

\begin{figure}[t]
\scalebox{0.28}[0.28]{\includegraphics{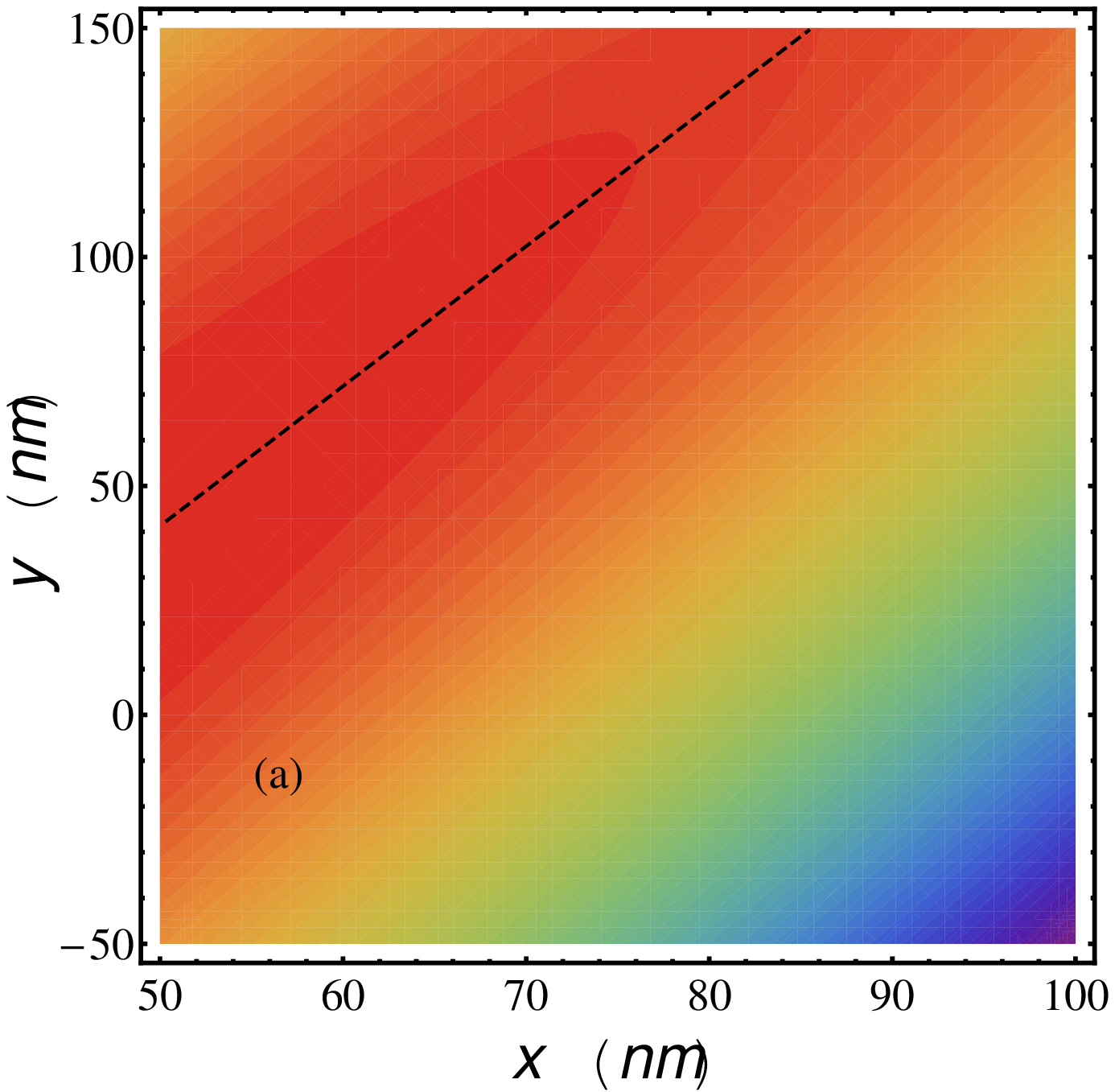}}
\scalebox{0.28}[0.28]{\includegraphics{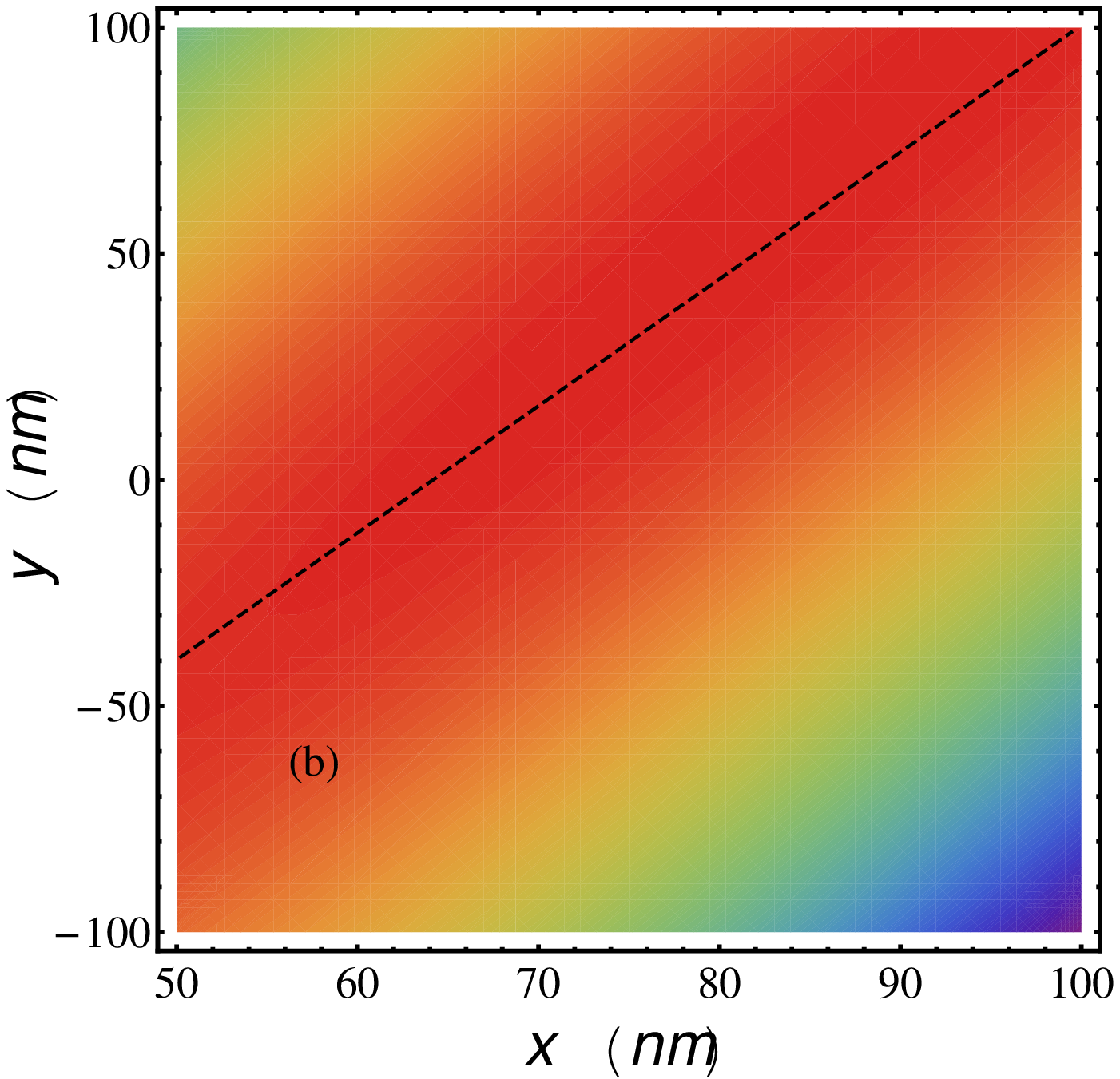}}
\scalebox{0.32}[0.32]{\includegraphics{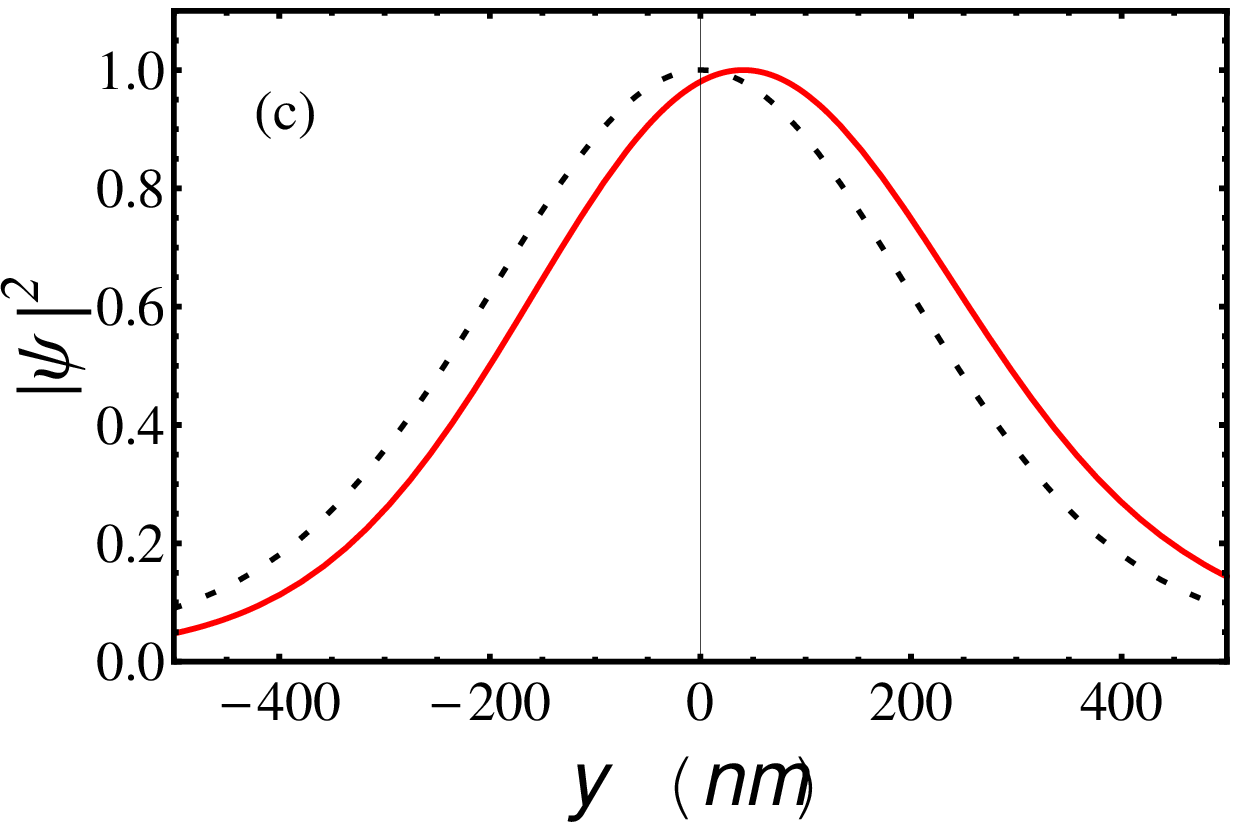}}
\scalebox{0.32}[0.32]{\includegraphics{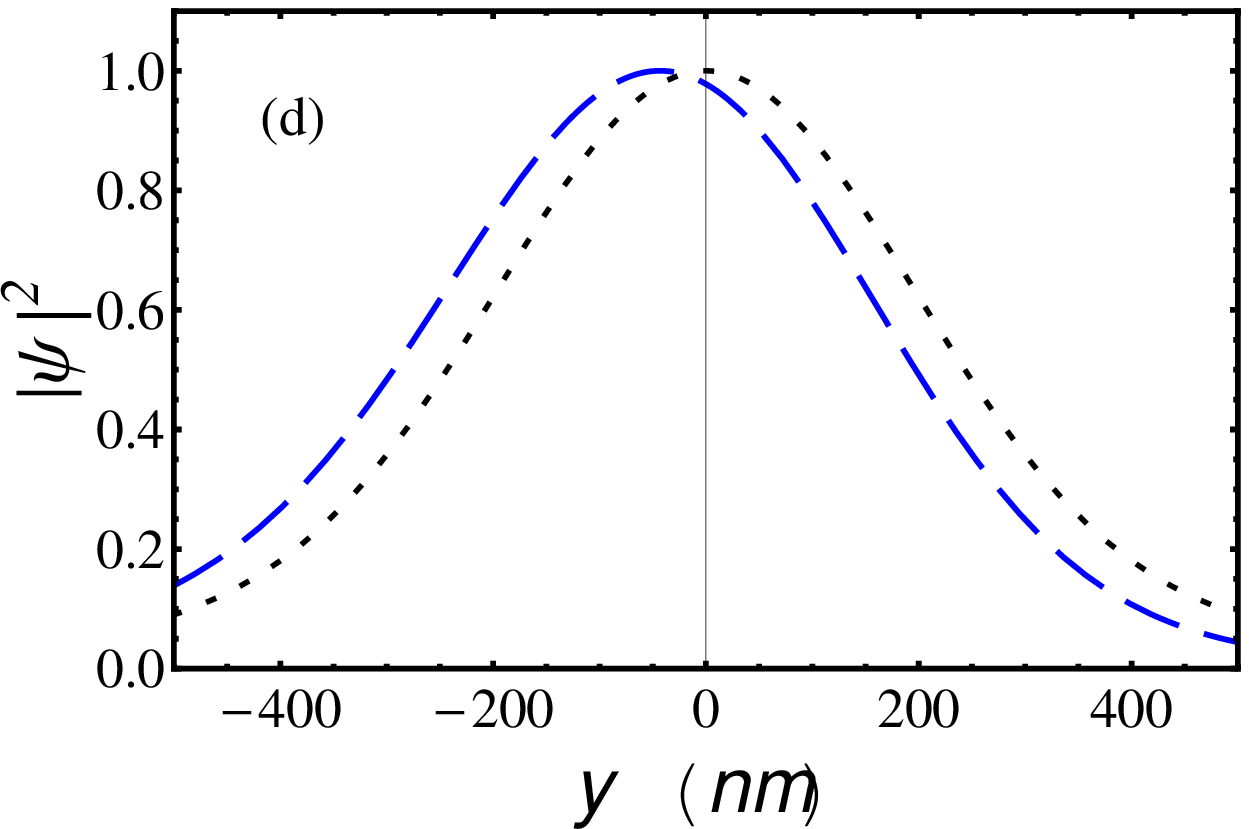}}
 \caption{(Color online) Distribution of the transmitted Gaussian-shaped beam for spin-up (a) and spin-down (b) polarized electrons,
 where $\theta_0=80^\circ$, beam width is $w=5 \lambda_e$,
 and the other physical parameters are the same as those in Fig. \ref{Fig.2}. Dashed lines represent the loci of maximum values for the eye.
 Profiles of the normalized incident and transmitted beams for spin-up (c) and spin-down (d) polarized electrons are also compared at $x=d$, where
 solid (red) and dashed (blue) lines
 represent the transmitted beams, and dotted (black) line represents the incident beam as reference. } \label{Fig.5}
\end{figure}

In this section, we further process to make the numerical simulations of the incident Gaussian-shaped beam,
$\Psi_{in}(x=0,y)= \exp(-y^2/2w^2_y+ik_{y0}y)$, where $w_y=w \sec\theta_0$, $w$ is the width of beam.
According to Fourier integral, using Eqs. (\ref{incident beam}) and (\ref{transmitted beam}),
we can obtain the wave function of transmitted beam as follows,
\begin{equation}
 \Psi_{tr} (x, y)= \frac{1}{\sqrt{2\pi}} \int_{-k}^{k} D A(k_y-k_{y0}) e^{i [k_x (x-d)+k_y y)]} dk_y,
\end{equation}
where $A(k_y-k_{y0})=w_y \exp[-(w^2_y/2)(k_y-k_{y0})^2]$.
For a well-collimated the range of above integral can be ideally extended from $- \infty$ to $\infty$.
Figs. \ref{Fig.5} (a) and (b) illustrate the distribution of transmitted beam, $ |\Psi_{tr} (x, y)|^2$,
for spin-up and spin-down polarized electrons, where $\theta_0=80^\circ$, the beam width is $w=5 \lambda_e$,
and the other physical parameters are the same as those in Fig. \ref{Fig.2}.
As demonstrated in Fig. \ref{Fig.5} (a) and (b), dashed lines represent the loci of the maximum values
for distribution of the transmitted beams for the eye. Evidently, the comparison between
the normalized incident and transmitted beams at $x=d$ in Fig. \ref{Fig.5} (c) and (d) show that the
positive and negative lateral shifts correspond to the different spin polarization. To find the numerical value of beam shifts, we can find the position $y$ corresponds to the maximum value of $ |\Psi_{tr} (x, y)|^2$. For the given
parameters in Fig. \ref{Fig.5}, $y^{+} = 40.80$ nm and $y^{-} = -43.21$ nm can be achieved for spin-up and spin-down
polarized electrons, which are large enough to sperate the different
spin-polarized electrons beam spatially.

\begin{figure}[t]
\scalebox{0.5}[0.5]{\includegraphics{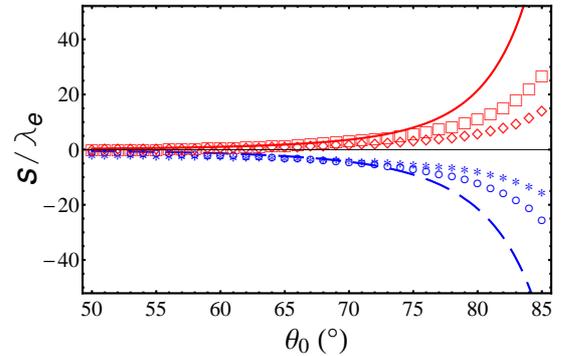}}
 \caption{(Color online) Influence of beam width on the Goos-H\"{a}nchen shifts,  where the physical parameters are the same as those in Fig. \ref{Fig.4}. Solid (red) and
 dashed (blue) lines represent the shifts for spin-up and spin-down polarized electrons, calculated by stationary phase method.
 Lines with markers ``$\square$" (``$\circ$") and ``$\diamond$" (``$\ast$") correspond to
 the numerical results for spin-up (spin-down) polarization with $w=10 \lambda_e$ and $w=5 \lambda_e$, respectively.} \label{Fig.6}
\end{figure}

In Fig. \ref{Fig.6}, the numerical results further show the influence of beam width on Goos-H\"{a}nchen shifts and spin beam splitter,
where the physical parameters are the same as those in Fig. \ref{Fig.4} and the beam width is chosen to be
$w=5 \lambda_e$ and $w=10 \lambda_e$ (corresponding to the beam divergence of $\delta \theta=4^\circ$ and $\delta \theta=2^\circ$). It is demonstrated that the wider the
local waist of incident beam is,  the closer to the theoretical results predicted by stationary phase method the
numerical results are.  It is worthwhile to point out
that there is discrepancy between theoretical results predicted by stationary phase approximation
and numerical results, resulting from the distortion of the transmitted electron beam \cite{XChen-PLA,Chen-PRB},
especially when the local beam waist is narrow, which implies large beam divergent $\delta \theta = \lambda_e/ (\pi w)$.
Actually, the stationary phase approximation is valid only for well-collimated electron beam \cite{XChen-PLA,Chen-PRB}.
In detail, when the incidence angle is small enough, the Goos-H\"{a}nchen shifts predicted by stationary phase method
are in agreement with the results given by numerical simulations. But when a large incidence angle is chosen,
$\delta \theta \ll 90^\circ-\theta_0$ is not satisfied, so that
the transmitted beam shape will undergo severe distortion.
It is suggested that one cannot achieve an arbitrary large Goos-H\"{a}nchen shift in practice by increasing the incidence angle,
taking into account the influence of finite beam width.
In a word, the theoretical results of stationary phase method and numerical simulations show that
the simultaneously large and opposite beam shifts for spin-up and spin-down polarized electrons
allow this system to realize the spin beam splitter, which can completely
separate spin-up and spin-down polarized electrons in such semiconductor quantum device.

\section{Discussion and Conclusion}

We have investigated the Goos-H\"{a}nchen shift for ballistic electrons
through a parabolic quantum well under a uniform magnetic field. Due to the effect of magnetic field, the Goos-H\"{a}nchen shift for ballistic electrons
also becomes spin-dependent, which leads to the novel applications in designing the spin filter
and spin beam splitter. It is found that
the Goos-H\"{a}nchen shift can be modulated by the incident energy, magnetic field, Landau quantum number,
and incidence angle. In addition, we would like to mention that the Goos-H\"{a}nchen shift is also sensitive to the device geometry,
since the lateral shift depends periodically on the width $d$ of parabolic quantum well region.
The spin-dependent Goos-H\"{a}nchen shift and spin beam splitter presented here can be discussed in GaAs/AlGaAs-based quantum well structure \cite{Wan},
rather than InSb semiconductor with high $g$-factor.

Last but not least, the Goos-H\"{a}nchen shift and relevant transverse Imbert-Fedorov effect in
such quantum well under the magnetic field may have close relation to spin Hall effect \cite{Sinitsyn,Onoda,Bliokh}, which
deserves further investigations. We hope all the results presented here will stimulate further theoretical and experimental researches
on quantum Goos-H\"{a}nchen shifts for ballistic electrons in semiconductor \cite{Chen-PRB,XChen-JAP} and
graphene \cite{Beenakker-PRL,Wu} and their applications in various quantum electronic devices.

\section*{Acknowledgments}
X. C. thanks Y. Ban and Z.-F. Zhang for helpful discussions.
This work was supported by the National Natural Science Foundation of
China (Grant Nos. 60806041 and 60877055) and the Shanghai Leading Academic Discipline
Program (Grant No. S30105). X. C. also acknowledges
funding by Juan de la Cierva Programme, Basque Government
(Grant No. IT472-10) and Ministerio de Ciencia e Innovaci\'on (FIS2009-12773-C02-01).


\end{document}